# Thermal expansion coefficient and lattice anharmonicity of cubic boron arsenide

Xi Chen[1], Chunhua Li[2], Fei Tian[3], Geethal Amila Gamage[3], Sean Sullivan[4], Jianshi Zhou[1,4], David Broido[2], Zhifeng Ren[3], and Li Shi[1,4]

[1]Materials Science and Engineering Program, Texas Materials Institute, The University of Texas at Austin, Austin, TX 78712, USA

[2]Department of Physics, Boston College, Chestnut Hill, MA 02467, USA

[3]Department of Physics and the Texas Center for Superconductivity, University of Houston, Houston, TX 77204, USA

[4]Department of Mechanical Engineering, The University of Texas at Austin, Austin, TX 78712, USA

E-mail: lishi@mail.utexas.edu

**Abstract**

Recent measurements of an unusual high thermal conductivity of around 1000 W m$^{-1}$ K$^{-1}$ at room temperature in cubic boron arsenide (BAs) confirm predictions from theory and suggest potential applications of this semiconductor compound for thermal management applications. Knowledge of the thermal expansion coefficient and Grüneisen parameter of a material contributes both to the fundamental understanding of its lattice anharmonicity and to assessing its utility as a thermal-management material. However, previous theoretical calculations of the thermal expansion coefficient and Grüneisen parameter of BAs yield inconsistent results. Here we report the linear thermal expansion coefficient of BAs obtained from the X-ray diffraction measurements from 300 K to 773 K. The measurement results are in good agreement with our *ab initio* calculations that account for atomic interactions up to fifth nearest neighbours. With the measured thermal expansion coefficient and specific heat, a Grüneisen parameter of BAs of 0.84±0.09 is obtained at 300 K, in excellent agreement with the value of 0.82 calculated from first principles and much lower than prior theoretical results. Our results confirm that BAs exhibits a better thermal expansion coefficient match with commonly used semiconductors than other high-thermal conductivity materials such as diamond and cubic boron nitride.

# 1. Introduction

Owing to the shrinking size and increasing density of transistors, heat dissipation has become a critical challenge for microelectronic devices [1]. Inefficient heat removal from localized hot spots in semiconductor devices compromises device performance and accelerates electromigration and thermomechanical failures [2,3]. One approach to addressing this challenge is to identify new semiconducting materials with a much higher thermal conductivity ($\kappa$) than existing values in current semiconductor devices. The highest known room-temperature $\kappa$ values, around 2000 W m$^{-1}$ K$^{-1}$, have been found in diamond and graphite [4-6]. However, diamond is an electrical insulator, and graphite is a semi-metal. In addition, the $\kappa$ of graphite is highly anisotropic with a low cross-plane thermal conductivity of only about 10 W m$^{-1}$ K$^{-1}$ [5]. Meanwhile, high-quality diamond is expensive and has not been synthesized on a large scale. An additional key problem with diamond is the considerable mismatch between its thermal expansion coefficient ($\alpha$) and the larger $\alpha$ values of common semiconductors such as silicon (Si) and gallium arsenide (GaAs). This mismatch can lead to thermally induced failure of the bonding between a semiconductor device and a diamond heat spreading layer.

First-principles calculations have predicted that cubic boron arsenide (BAs) could have a high $\kappa$ comparable to that of diamond and graphite [7,8]. The theory has motivated experimental efforts to synthesize and measure thermal transport in BAs crystals [9-12]. Although the presence of impurities [13] and defects [14] has been shown to reduce the $\kappa$ of the BAs crystals, high $\kappa$ values of around 1000 W m$^{-1}$ K$^{-1}$ at room temperature have recently been measured in BAs crystals grown using a chemical vapour transport (CVT) method [15-17]. In addition, the measured $\kappa$ exhibits strong temperature dependence that reveals an unusually important role of four-phonon scattering in the thermal transport of BAs [15].

Along with the thermal conductivity, the thermal expansion coefficient and Grüneisen parameter ($\gamma$) are two other important properties that contribute to fundamental understanding of the lattice anharmonicity and help assess the utility of a material in thermal-management applications. Prior theoretical calculations have obtained somewhat different $\alpha$ values for BAs. A molecular dynamics (MD) simulation of BAs based on a three-body Tersoff potential has obtained a room-temperature linear thermal expansion coefficient ($\alpha_l$) of 4.1 x 10$^{-6}$ K$^{-1}$ [18]. In comparison, first-principles calculations of BAs have yielded a different $\alpha_l$ value of 3.04 x 10$^{-6}$ K$^{-1}$ at 300 K [19]. Recently, a large volumetric thermal expansion coefficient ($\alpha_v$) of 3.27 x 10$^{-5}$ K$^{-1}$ corresponding to a $\alpha_l$ of 10.9 x 10$^{-6}$ K$^{-1}$ at 300 K for BAs has been obtained by another first-principles method [20]. However, there has been no experimental report of the thermal expansion coefficient of BAs. In addition, three existing theoretical calculations of BAs have obtained $\gamma$ values based on different models and approximations [20-22]. Because the Grüneisen parameter characterizes the lattice anharmonicity, it is necessary to investigate this important fundamental property further via both experiments and rigorous theoretical calculations.

Here we report both experimental and theoretical studies of thermal expansion coefficients, specific heat ($C_p$), and the Grüneisen parameter of BAs. The linear thermal expansion coefficient from 300 K to 773 K is determined by high-temperature X-ray diffraction (XRD) measurements. The room temperature $\alpha_l$ is found to be (4.2 ± 0.4) x 10$^{-6}$ K$^{-1}$ at 300 K and to increase with increasing temperature. We find good agreement between our measured data and our first-principles calculations over the full temperature range considered when interatomic interactions up to fifth neighbours are accounted for in the calculations. The $\alpha$ values of BAs makes the compound a much better match to silicon and other III-V semiconductor compounds than diamond, and gives insight into the phonon and anharmonic properties of BAs. Together with the separately measured bulk modulus, the measured thermal expansion coefficient and specific heat data are used to obtain the Grüneisen parameter as 0.84±0.09 at room temperature, in very good agreement with our first-principles calculation result of 0.82 and much smaller than two prior calculation results [20,21].

## 2. Experimental and Theoretical Methods

The single crystals of BAs were grown by a CVT method in a sealed quartz tube [15]. The XRD patterns of BAs at different temperatures were acquired using a Scintag X1 Theta-Theta Diffractometer with a Cu K$\alpha$ radiation source. For the XRD study, several BAs crystals were ground into powder. The $C_p$ of BAs was measured using a Physical Properties Measurement System (Quantum Design).

In order to provide theoretical values for comparison with the measurement results, we have performed first-principles calculations of the thermal expansion coefficient of BAs, which can be defined in the quasi-harmonic approximation as [23]

$$\alpha_l = \frac{1}{3B_T}\sum_\lambda C_\lambda \gamma_\lambda \qquad (1)$$

where $B_T$ is the bulk modulus of BAs, the sum is over phonon modes $\lambda$, and $C_\lambda$ and $\gamma_\lambda$ are, respectively, the mode specific heat and mode Grüneisen parameter. Here we calculate $\gamma_\lambda$ as:

$$\gamma_\lambda = -\frac{V}{\omega_\lambda}\frac{d\omega_\lambda}{dV} = -\frac{1}{6\omega_\lambda^2}\sum_k \sum_{l'k'}\frac{\varepsilon_{\alpha k}^{\lambda*}\varepsilon_{\beta k'}^{\lambda}}{\sqrt{M_k M_{k'}}}e^{i\mathbf{q}\cdot\mathbf{R}_{l'}}\xi_{\alpha\beta}(k,l'k') \qquad (2)$$

with $\xi_{\alpha\beta}(k,l'k') = \sum_{l''k''}\Phi_{\alpha\beta\gamma}(0k,l'k',l''k'')r_{l''k''\gamma}$. In the equations, $V$ is the volume, $\omega_\lambda$ is the phonon angular frequency, $\alpha,\beta,\gamma$ are Cartesian components, $lk$ labels the $k^{th}$ atom in the $l^{th}$ unit cell, $\varepsilon_{\alpha k}^{\lambda}$ is the $\alpha^{th}$ component of the phonon eigenvector, $M_k$ is the isotope averaged mass of atom $k$, $R_l$ is a lattice vector, and $\mathbf{r}_{lk}$ is a vector locating the $k^{th}$ atom in the $l^{th}$ unit cell. The terms $\Phi_{\alpha\beta\gamma}(0k,l'k',l'',k'')$ are the third-order anharmonic interatomic force constants (IFCs). The details of their calculation and those of the phonon frequencies are similar to those described in Ref. 19. Calculations were performed within the framework of density functional theory with norm-conserving pseudopotentials in the local density approximation (LDA). The calculated phonon dispersions of BAs at different temperatures show only a slight softening of optic modes [24].

As a check of the calculated results, Figure 1 shows the BAs mode Grüneisen parameters plotted along high symmetry directions. This figure shows results from both frequency derivative approach according to the first expression of Eq. (2) and the third-order interatomic force constant approach based on the second expression in Eq. (2). The two sets of results are in good agreement with some small deviations seen in the lowest transverse acoustic branches. As discussed below, the thermal expansion coefficients calculated using the two expressions in Eq. (2) are almost the same.

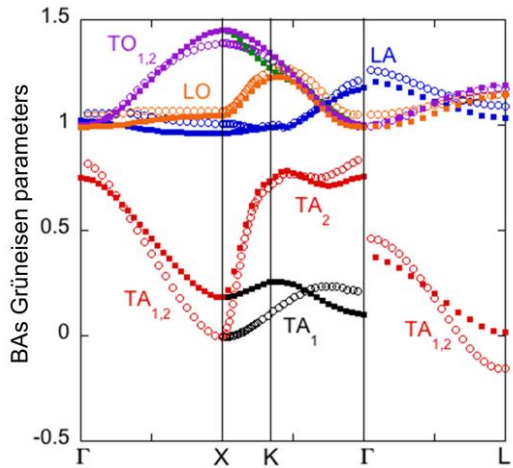

**Figure 1**. The calculated mode Grüneisen parameters of BAs. The open and solid symbols are determined using the frequency derivative approach and third-order interatomic force constant approach, respectively.

## 3. Results and Discussion

BAs exhibits a cubic zinc blende crystal structure with a space group of $F\bar{4}3m$, as shown in Figure 2(a). Figure 2(b) shows a photo of a typical BAs crystal grown by the CVT method [15]. Similar to some other BAs crystals, the sample exhibits a semitransparent reddish colour and plate-like shape with a lateral dimension on the order of 2-3 millimeters that are adequate for this study while that larger crystals have also been grown.

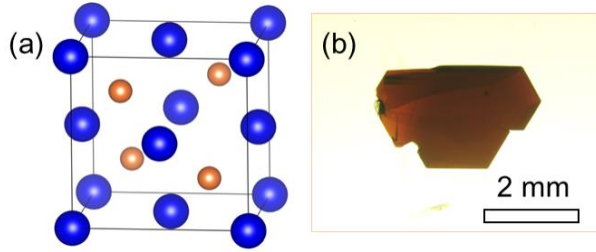

**Figure 2.** (a) The crystal structure of BAs. The blue and yellow spheres represent As and B atoms, respectively. (b) A photo of a BAs crystal grown by the CVT method.

Figure 3 shows the temperature dependence of the $C_p$ of BAs. At low temperatures, the $C_p$ of semiconducting BAs consists of contributions from phonons and electrons as

$$C_p(T) = \frac{12xN_A \pi^4 k_B}{5M}\left(\frac{T}{\theta_D}\right)^3 + bT \qquad (3)$$

where $N_A$ is Avogadro's constant, $M$ is the molar mass, $T$ is temperature, $k_B$ is Boltzmann constant, $\theta_D$ is the Debye temperature, $x$ is the number of atoms per formula unit, and $b$ is the Sommerfeld coefficient. The fitting of low-temperature $C_p$ according to Eq. (3) is shown in the inset of Fig. 3. The obtained $\theta_D$ is about 668 K, which is consistent with the value of 700 K obtained from the first-principles calculation [7]. In addition, the obtained $b$ is about $6.5 \times 10^{-7}$ J g$^{-1}$ K$^{-2}$, which is due to the semiconducting nature of BAs.

The specific heat at constant pressure ($C_p$) and constant volume ($C_v$) of BAs are calculated by the first-principles method. The calculated values agree well with the experimental data and approach the Dulong-Petit limit at high temperature, as shown in Fig. 3. It should be noted that the $C_p$ value is significantly higher than $C_v$ above 400 K as a result of thermal expansion.

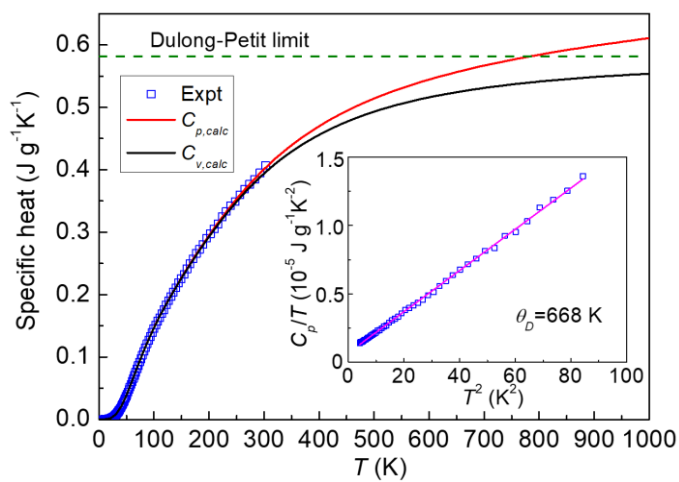

**Figure 3.** Temperature dependence of measured $C_p$ for BAs in comparison with the calculated $C_v$ and $C_p$. The inset is a plot of $C_p(T)/T$ vs $T^2$. The pink line in the inset is the fitting curve according to Eq. (3).

Figure 4 shows the obtained XRD patterns of BAs in air for temperatures between 300 K to 773 K. We found that the BAs sample is stable up to 773 K since no phase change was observed. The observation is consistent with a previous study, in which cubic BAs was reported to be stable up to 1193 K [25]. It should be noted that the XRD intensity of BAs is reduced with increasing temperature, which could be due to the enhanced lattice vibrations [26].

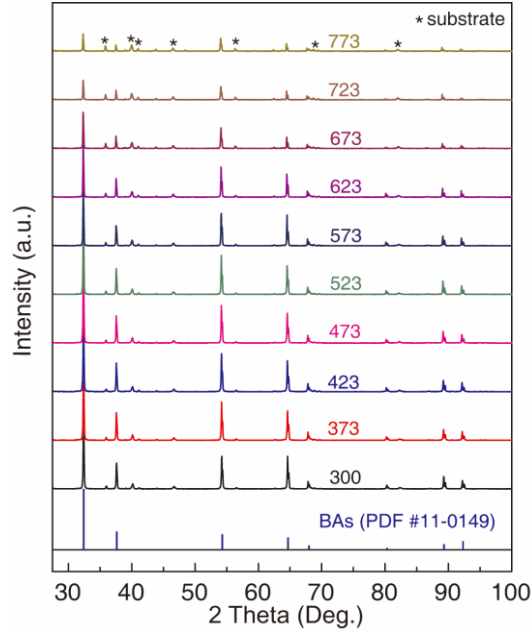

**Figure 4.** The XRD patterns of BAs measured at different temperatures.

The lattice parameter $a$ of BAs is calculated according to Bragg's law [27],

$$n\lambda_{\text{x-ray}} = 2d_{hk\ell} \sin\theta, \quad (4)$$

where $n$ is the order of diffraction; $\lambda_{\text{x-ray}}$ is the X-ray wavelength; $d$ is the inter-plane distance; $\theta$ is the Bragg's angle; and $h$, $k$, and $\ell$ are the Miller indices. The inter-plane distance $d$ is related to the lattice constant $a$ of the cubic crystal system as $d = \frac{a}{\sqrt{h^2+k^2+\ell^2}}$. The lattice parameter is obtained using the Jade software [28] with displacement correction. Figure 5(a) shows the temperature dependence of the lattice parameter for BAs. The value of $a$ is $4.7785 \pm 0.0002$ Å at 300 K, in good agreement with previous reports [9,11,29]. The $a$ value increases monotonically with temperature. The temperature dependence of $a$ can be fitted by a second-order polynomial as [30]

$$a(T) = a_0 + a_1 T + a_2 T^2, \quad (5)$$

where $a_0$, $a_1$ and $a_2$ are fitting parameters. Fitting the measurement data with a third- or fourth-order polynomial results in increasing fitting errors. The obtained values are $a_0 = 4.7735$ Å, $a_1 = 1.29141 \times 10^{-5}$ Å K$^{-1}$, and $a_2 = 1.22755 \times 10^{-8}$ Å K$^{-2}$.

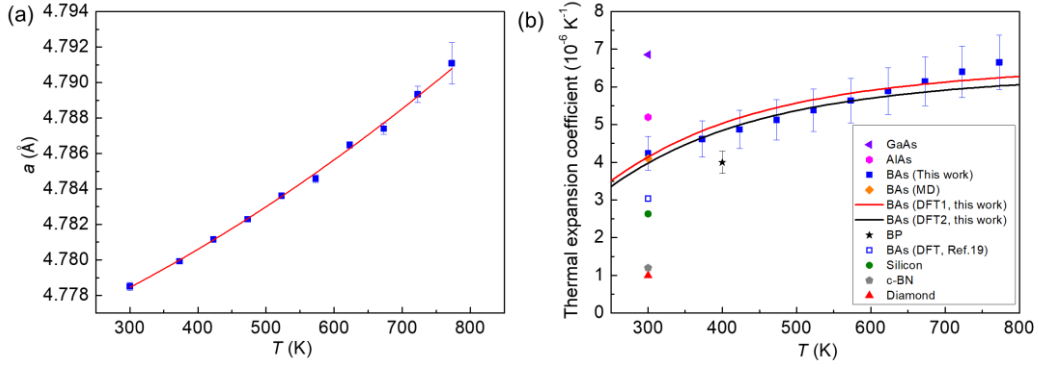

**Figure 5.** (a) The lattice parameter $a$ of BAs at different temperatures. The red line is the polynomial fitting according to Eq. (5). (b) The measured $\alpha_l$ of BAs at different temperatures. Shown for comparison are the calculated $\alpha_l$ values of BAs by the MD method [18] and the density functional theory (DFT) method [19] and the reported measured linear thermal expansion coefficient values of GaAs [31], AlAs [32], BP [33], c-BN [34], silicon [35], and diamond [36]. The red and black lines in (b) are the calculated $\alpha_l$ using the first and second expression in Eq. (2), respectively.

According to the definition of the linear thermal expansion coefficient $\alpha_l$, $\alpha_l \equiv \frac{1}{a}\frac{da}{dT}$, we obtain the temperature dependence of the thermal expansion coefficient as,

$$\alpha_l(T)=(a_1+2a_2T)/(a_0+a_1T+a_2T^2). \qquad (6)$$

Due to the second-order polynomial fitting, the obtained $\alpha_l$ of BAs increases linearly with temperature from $(4.2\pm0.4) \times 10^{-6}$ K$^{-1}$ at 300 K to $(6.7\pm0.7) \times 10^{-6}$ K$^{-1}$ at 773 K. The $\alpha_v$ of cubic BAs can be approximated as $\alpha_v = 3\alpha_l$, which gives about $1.26 \times 10^{-5}$ K$^{-1}$ at 300 K.

In comparison, the present first-principles calculations using the second expression in Eq. (2) obtains $\alpha_l = 4.0 \times 10^{-6}$ K$^{-1}$ at 300 K, considerably larger than the value of $\alpha_l = 3.04 \times 10^{-6}$ K$^{-1}$ calculated in Ref. 19. Using the same IFCs as used in Ref. 19, we have established that this difference is primarily due to the longer range of the third-order IFCs used in the present calculation, extending to the fifth-nearest neighbour here, as compared to only the third-nearest neighbour in Ref. 19. In addition, the bulk modulus of $B_T = 1.46$ Mbar obtained here is consistent with several other previously reported values calculated within the LDA [37-39] and a recent measurement [40], but smaller than the 1.57 Mbar determined in Ref. 19. A small (~5%) error also occurred in Ref. 19 due to the calculations being performed for a rock salt rather than a zinc blende structure. It is worth noting that a similar room-temperature $\alpha_l$ value of $4.1 \times 10^{-6}$ K$^{-1}$ was obtained from a MD calculation using empirical potentials [18].

Figure 5 (b) compares the measured $\alpha$ to the calculated values using the two expressions in Eq. 2. The BAs $\alpha$ curves from the present first-principles calculations are in excellent agreement with the measured data for temperatures below about 600 K, beyond which the measurement data becomes slightly higher than the calculated curves. This small discrepancy at high temperatures may be due to either higher-order anharmonic effects that are not captured with the theoretical approach used here, or the second-order polynomial fitting of the measured lattice parameter data. It should be noted that the calculated $\alpha_l$ using the frequency derivative method according to the first expression in Eq. (2) is very close to the values obtained using the third order force constant approach according to the second expression in Eq. (2) with the former being only about 4% higher than the latter above 300 K.

Also shown in Fig. 5 (b) is a comparison of the measured and calculated $\alpha_l$ data of BAs with other cubic-phase materials that have been commonly used or are being explored for electronic and optoelectronic devices. In this figure, the $\alpha_l$ values of both BAs and BP [33] lie between those of Si [35] and AlAs [32] or GaAs [31], while those of diamond [36] and cubic-phase BN [34] both lie well below the $\alpha_l$ values of these commonly used semiconductors. Specifically, BAs shows a ~50% higher and ~40% lower $\alpha_l$ value at room temperature than Si and GaAs, respectively. In comparison, the thermal expansion coefficient of diamond is almost three times smaller than that of Si and almost seven times smaller than that of GaAs. Thermal stress analysis indicates that BAs substrate can lead to a much smaller thermal stress for a GaAs thin film device than both diamond and c-BN substrates [24].

The measurement results allow us to determine the mode averaged Grüneisen parameter of BAs. The macroscopic formulation of $\gamma$ is given by [41]

$$\gamma = \frac{\alpha_v B_T}{C_v \rho}, \qquad (7)$$

where $\rho$ is the density of the sample. With the measured specific heat and $\alpha_v$ in this work as well as the $B_T$ obtained from a separate measurement [40], the $\gamma$ of BAs is calculated to be 0.84±0.09 at 300 K. It should be noted that the measured $C_p$ is used in the calculation as the calculated difference between $C_p$ and $C_v$ is small at 300 K. Using the calculated value of $\alpha_l$ from Eq. (1) along with the calculated $C_v$ and $B_T$ gives a $\gamma$ value of 0.82, in excellent agreement with the value obtained from the measured data. In comparison, a recent density-functional perturbation theory simulation has obtained the $\gamma$ of about 1 for the phonon mode at about 700 cm$^{-1}$ [22]. It should be noted that the obtained $\gamma$ of BAs in this work from both measured data and first-principles calculations is less than half of the value calculated by a first-principles approach that extracts specific heat and Grüneisen parameters employing a Debye approximation and without explicitly calculating the per mode contributions [20], and about half that obtained using a shell model [21]. This highlights the importance of performing fully first principles calculations to determine thermal properties of materials.

## 4. Conclusions

In summary, we have investigated the specific heat, thermal expansion coefficient and Grüneisen parameter of BAs both experimentally and theoretically. The measured thermal expansion coefficient value is higher than the previous first-principles calculation results that used short-ranged anharmonic force constants [19]. It is found that extension up to fifth-nearest neighbour is necessary for the first-principles calculation to obtain results in agreement with the measurements. With this extension, the calculated room-temperature $\alpha_l$, $4.0 \times 10^{-6}$ K$^{-1}$, is close to the corresponding measured value, $(4.2\pm0.4) \times 10^{-6}$ K$^{-1}$, and the difference between the calculation result and measurement is within the measurement uncertainty over the temperature range between 300 K and 773 K. Based on this comparison between the measurement and calculation results, long-range atomic interaction plays an important role in the lattice anharmonicity of BAs. Importantly, the results validate that cubic-phase BAs has a much smaller thermal expansion mismatch to common semiconductors as compared to diamond and cubic boron nitride. This reduced mismatch is a desirable attribute for the integration of BAs in existing silicon and other III-V semiconductor device architectures. Together with the separately measured bulk modulus [40], the specific heat and thermal expansion coefficient measured here allows us to determine a room-temperature value of $0.84\pm0.09$ for the Grüneisen parameter, which is in excellent agreement with the corresponding value of 0.82 calculated from first principles and much smaller than previous theoretical results. These findings give fundamental insight into the thermal properties of BAs and confirm it as a technologically promising material for thermal management applications.

## Acknowledgements

The work is supported by the Office of Naval Research under a MURI grant N00014-16-1-2436. The authors thank Steve Swinnea and Texas Materials Institute for assistance with the high-temperature XRD measurements.

**References**

[1] A. L. Moore and L. Shi, Emerging challenges and materials for thermal management of electronics, Mater. Today **17**, 163 (2014).

[2] H. Yasunaga and A. Natori, Electromigration on semiconductor surfaces, Surf. Sci. Rep. **15**, 205 (1992).

[3] M. Ciappa, Selected failure mechanisms of modern power modules, Microelectron. Reliab. **42**, 653 (2002).

[4] R. Berman, E. L. Foster, J. M. Ziman, and F. E. Simon, The thermal conductivity of dielectric crystals: the effect of isotopes, Proc. R. Soc. Lond. A. Math. Phys. Sci. **237**, 344 (1956).

[5] Thermophysical Properties Research Center, Purdue University, Thermophysical Properties of Matter (IFI, 1970 -1979).

[6] G. A. Slack, Nonmetallic crystals with high thermal conductivity, J. Phys. Chem. Solids **34**, 321 (1973).

[7] L. Lindsay, D. A. Broido, and T. L. Reinecke, First-Principles Determination of Ultrahigh Thermal Conductivity of Boron Arsenide: A Competitor for Diamond?, Phys. Rev. Lett. **111**, 025901 (2013).

[8] T. Feng, L. Lindsay, and X. Ruan, Four-phonon scattering significantly reduces intrinsic thermal conductivity of solids, Phys. Rev. B **96**, 161201 (2017).

[9] B. Lv *et al.*, Experimental study of the proposed super-thermal-conductor: BAs, Appl. Phys. Lett. **106**, 074105 (2015).

[10] J. Kim, D. A. Evans, D. P. Sellan, O. M. Williams, E. Ou, A. H. Cowley, and L. Shi, Thermal and thermoelectric transport measurements of an individual boron arsenide microstructure, Appl. Phys. Lett. **108**, 201905 (2016).

[11] J. Xing, X. Chen, Y. Zhou, J. C. Culbertson, J. A. FreitasJr., E. R. Glaser, J. Zhou, L. Shi, and N. Ni, Multimillimeter-sized cubic boron arsenide grown by chemical vapor transport via a tellurium tetraiodide transport agent, Appl. Phys. Lett. **112**, 261901 (2018).

[12] F. Tian *et al.*, Seeded growth of boron arsenide single crystals with high thermal conductivity, Appl. Phys. Lett. **112**, 031903 (2018).

[13] J. L. Lyons *et al.*, Impurity-derived p-type conductivity in cubic boron arsenide, Appl. Phys. Lett. **113**, 251902 (2018).

[14] Q. Zheng, C. A. Polanco, M.-H. Du, L. R. Lindsay, M. Chi, J. Yan, and B. C. Sales, Antisite Pairs Suppress the Thermal Conductivity of BAs, Phys. Rev. Lett. **121**, 105901 (2018).

[15] F. Tian *et al.*, Unusual high thermal conductivity in boron arsenide bulk crystals, Science **361**, 582 (2018).

[16] S. Li, Q. Zheng, Y. Lv, X. Liu, X. Wang, P. Y. Huang, D. G. Cahill, and B. Lv, High thermal conductivity in cubic boron arsenide crystals, Science **361**, 579 (2018).

[17] J. S. Kang, M. Li, H. Wu, H. Nguyen, and Y. Hu, Experimental observation of high thermal conductivity in boron arsenide, Science **361**, 575 (2018).

[18] F. Benkabou, C. Chikr.Z, H. Aourag, P. J. Becker, and M. Certier, Atomistic study of zinc-blende BAs from molecular dynamics, Phys. Lett. A **252**, 71 (1999).

[19] D. A. Broido, L. Lindsay, and T. L. Reinecke, Ab initio study of the unusual thermal transport properties of boron arsenide and related materials, Phys. Rev. B **88**, 214303 (2013).

[20] S. Daoud, N. Bioud, and N. Lebga, Elastic and thermophysical properties of BAs under high pressure and temperature, Chin. J. Phys. **57**, 165 (2019).

[21] D. Varshney, G. Joshi, M. Varshney, and S. Shriya, Pressure induced mechanical properties of boron based pnictides, Solid State Sci. **12**, 864 (2010).

[22] V. G. Hadjiev, M. N. Iliev, B. Lv, Z. F. Ren, and C. W. Chu, Anomalous vibrational properties of cubic boron arsenide, Phys. Rev. B **89**, 024308 (2014).

[23] T. H. K. Barron and M. L. Klein, in Dynamical Properties of Solids, edited by G. K. Horton and A. A. Maradudin (North-Holland, Amsterdam, 1974), Vol. I, p. 391.

[24] See Supplemental Material for the calculated phonon dispersions of BAs at different temperatures and thermal stress analysis.

[25] T. L. Chu and A. E. Hyslop, Preparation and properties of boron arsenide films, J. Electrochem. Soc. **121**, 412 (1974).

[26] M. Saleem and D. Varshney, Structural, thermal, and transport properties of $La_{0.67}Sr_{0.33}MnO_3$ nanoparticles synthesized via the sol–gel auto-combustion technique, RSC Adv. **8**, 1600 (2018).

[27] A. Zamkovskaya, E. Maksimova, I. Nauhatsky, and M. Shapoval, X-ray diffraction investigations of the thermal expansion of iron borate $FeBO_3$ crystals, J. Phys. Conf. Ser. **929**, 012030 (2017).

[28] Jade version 9.1, Materials Data Inc., Livermore, CA.

[29] T. L. Chu and A. E. Hyslop, Crystal Growth and Properties of Boron Monoarsenide, J. Appl. Phys. **43**, 276 (1972).

[30] W. A. Paxton *et al.*, Anisotropic Thermal Expansion of Zirconium Diboride: An Energy-Dispersive X-Ray Diffraction Study, J. Ceram. **2016**, 5, 8346563 (2016).

[31] E. D. Pierron, D. L. Parker, and J. B. McNeely, Coefficient of Expansion of GaAs, GaP, and Ga(As, P) Compounds from −62° to 200°C, J. Appl. Phys. **38**, 4669 (1967).

[32] M. Ettenberg and R. J. Paff, Thermal Expansion of AlAs, J. Appl. Phys. **41**, 3926 (1970).

[33] M. Takashi, O. Jun, N. Tatau, and U. Susumu, Thermal Expansion Coefficient of Boron Monophophide, Jpn. J. Appl. Phys. **15**, 1305 (1976).

[34] G. A. Slack and S. F. Bartram, Thermal expansion of some diamondlike crystals, J. Appl. Phys. **46**, 89 (1975).

[35] H. Watanabe, N. Yamada, and M. Okaji, Linear Thermal Expansion Coefficient of Silicon from 293 to 1000 K, Int. J. Thermophys. **25**, 221 (2004).

[36] H. Katsuji, M. Hiroshi, O. Kazutoshi, and K. Takuro, Thermal Expansion Coefficient of Synthetic Diamond Single Crystal at Low Temperatures, Jpn. J. Appl. Phys. **31**, 2527 (1992).

[37] S. Bağci, S. Duman, H. M. Tütüncü, and G. P. Srivastava, Electronic and phonon properties of B$X$ (110) ($X$=P, As, and Sb) and Be$Y$ (110) ($Y$= S, Se, and Te) surfaces, Phys. Rev. B **79**, 125326 (2009).

[38] D. Touat, M. Ferhat, and A. Zaoui, Dynamical behaviour in the boron III–V group: a first-principles study, J. Phys. Condens. Matter **18**, 3647 (2006).

[39] B. Bouhafs, H. Aourag, and M. Certier, Trends in band-gap pressure coefficients in boron compounds BP, BAs, and BSb, J. Phys. Condens. Matter **12**, 5655 (2000).

[40] F. Tian *et al.*, Mechanical Properties of Boron Arsenide Single Crystal, Appl. Phys. Lett. **114**, 131903 (2019).

[41] N. L. Vočadlo and G. D. Price, The Grüneisen parameter — computer calculations via lattice dynamics, Phys. Earth Planet. Inter. **82**, 261 (1994).

# Supplemental Material

# Thermal expansion coefficient and lattice anharmonicity of cubic boron arsenide

Xi Chen[1], Chunhua Li[2], Fei Tian[3], Geethal Amila Gamage[3], Sean Sullivan[4], Jianshi Zhou[1,4], David Broido[2], Zhifeng Ren[3], and Li Shi[1,4]

[1]Materials Science and Engineering Program, Texas Materials Institute, The University of Texas at Austin, Austin, TX 78712, USA
[2]Department of Physics, Boston College, Chestnut Hill, MA 02467, USA
[3]Department of Physics and the Texas Center for Superconductivity, University of Houston, Houston, TX 77204, USA
[4]Department of Mechanical Engineering, The University of Texas at Austin, Austin, TX 78712, USA
E-mail: lishi@mail.utexas.edu

Figure S1 shows the phonon dispersions of cubic BAs from our first-principles calculations within the quasi-harmonic approximation at 200 K, 300 K and 600 K. The phonon dispersions are nearly identical for all three temperatures, showing only a slight softening of optic modes across this temperature range.

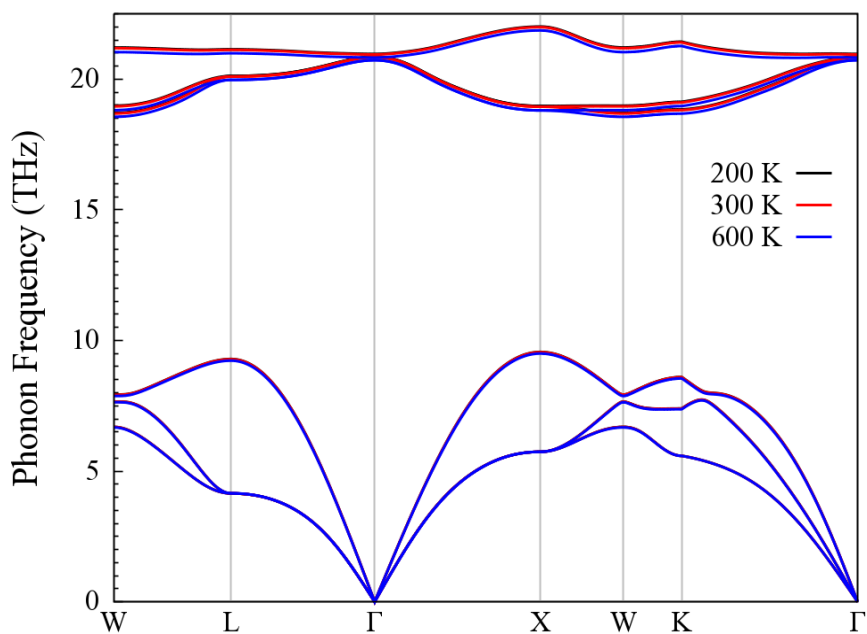

**Figure S1.** Phonon dispersions of cubic BAs calculated from first-principles calculations at different temperatures.

Thermal stresses (σ) are stresses induced in a body as a result of changes in temperature. Thermal stress in a thin film device can be calculated as

$$\sigma = E_A(\alpha_{l,B} - \alpha_{l,A})\Delta T, \qquad (S1)$$

where $E$ is Young's modulus, $\alpha_l$ is the linear thermal expansion coefficient, and $\Delta T$ is the temperature change. Figure S2 shows the calculated thermal stress for GaAs with other high thermal conductivity materials. It is found that the thermal stress is more than 50% smaller for GaAs/BAs than GaAs/diamond and GaAs/c-BN.

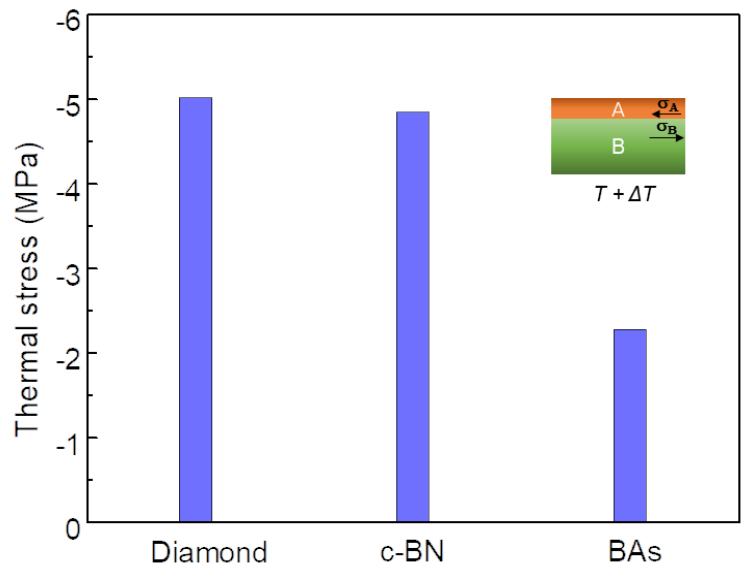

**Figure S2.** Calculated thermal stress in the thin film device. Material A is GaAs, and materials B is diamond, c-BN or BAs. The initial temperature is 300 K and the temperature rise is 10 K.